\documentclass[3p, twocolumn]{elsarticle}
\journal{Ultramicroscopy}
\usepackage{amsmath}
\usepackage{amssymb}
\usepackage{graphicx}
\usepackage{dcolumn}
\usepackage{bm}
\usepackage{gensymb} 
\usepackage[colorlinks=true,linkcolor=blue,anchorcolor=blue,citecolor=blue,urlcolor=blue]{hyperref}

\begin{document}

\begin{frontmatter}

\title{Temporal magnification for streaked ultrafast electron diffraction and microscopy}

\author[ucla]{D. Cesar}
\ead{dcesar@ucla.edu}

\author[ucla]{P. Musumeci}

\address[ucla]{Department of Physics and Astronomy, UCLA, Los Angeles, California 90095, USA}

\begin{abstract}
One of the frontiers of modern electron scattering instrumentation is improving temporal resolution in order to enable the observation of dynamical phenomena at their fundamental time-scales. We analyze how a radiofrequency cavity can be used as an electron longitudinal lens in order to produce a highly magnified temporal replica of an ultrafast process, and, in combination with a deflecting cavity, enable streaked electron images of optical-frequency phenomena. We present start-to-end simulations of an MeV electron beamline for two variations of this idea (a ``magnifying-glass'' and a ``point-projection'' configuration) showing the feasibility for an electron probe to achieve single shot 1.4\,fs(rms) temporal resolution.
\end{abstract}

\end{frontmatter}


\section{Introduction}
%
Time-resolved electron scattering (diffraction and microscopy) has proved to be a powerful tool for studying rapid microscopic changes in materials and uncovering new physical processes  \cite{zewail_4d_2006,dwyer_femtosecond_2006,king_ultrafast_2005,sciaini_femtosecond_2011}. Recently, the push to extend the reach of time-resolved electron instrumentation to investigate faster and more complex phenomena \cite{hall_future_2014} has fueled the development of a variety of ultrafast beam-based techniques with unique capabilities.

Most schemes operate in a conventional pump-probe modality in which a `pump' laser initiates a process and, after a carefully controlled delay, a short pulse of electrons `probes' the structure of the sample. The time evolution is recorded by varying the pump-probe delay while recording many still-frames. This approach requires the dynamics to unfold in the same way after each trigger; for even in the case that each frame is exposed in a single-shot, reconstructing the evolution of the system requires collating a series of independent events \cite{weathersby_mega-electron-volt_2015,van_oudheusden_compression_2010,musumeci_laser-induced_2010}. An alternative approach, analogous to a streak camera, allows us to record the entire evolution of a single event by using a deflecting cavity to streak the temporal distribution of a beam along one of the transverse coordinates \cite{Mourou&Williamson,li:continuousUED,musumeci_capturing_2010}. This typically requires the loss of one spatial dimension, and thus the use of a slit to select a row from the diffraction pattern, however the loss of information can be avoided if the electron current profile is not continuous but made of discrete pulses which can be weakly separated by a fast deflector\,\cite{musumeci_double-shot_2017} or if compressed sensing techniques are used to reconstruct the voxels\,\cite{qi_compressed_2018}.

In this streaking modality, the temporal resolution is set by the strength of the deflection element and so it can be much shorter than the electron pulse length or pump-probe jitter (which limits most pump-probe techniques). For example, single shot temporal resolution of 30\,fs (rms) has been demonstrated using a 9.6 GHz RF cavity and a 3 MeV electron beam to monitor the fast expansion of an electron cloud generated when an ultrashort laser pulse hits a metal surface\cite{scoby_single-shot_2013}. 

When streaking, the temporal resolution is ultimately limited by i) the emittance of the probing high-current e-beam and ii) by the intensity and frequency of the transverse deflecting field \cite{akre_transverse_2001}. The first one is set by the source brightness, while the second one is bound by the available power sources, or at the very high field limit by breakdown phenomena in the deflecting cavity itself. A variety of advanced techniques have been proposed to improve on this resolution, for example by implementing a complex electron optical setup to obtain longitudinal to spatial imaging \cite{xiang_longitudinal--transverse_2010} or by using THz-based deflecting structures \cite{fabianska_split_2014,zhao_terahertz_2018,li_terahertz-based_2018}.

An interesting approach to further improve the temporal resolution is offered by the realization that accelerating RF cavities can be used as longitudinal lenses to magnify temporal features in the beam. Such cavities are already installed in many beamlines where the analogy to a lens has been profitably used to describe the compression of chirped electron pulses \cite{van_oudheusden_compression_2010,chatelain_ultrafast_2012,gao_full_2012,maxson_direct_2017} and timing jitter reduction schemes \cite{franssen_improving_2017}.

\begin{figure}
\centering
\includegraphics[width=\linewidth]{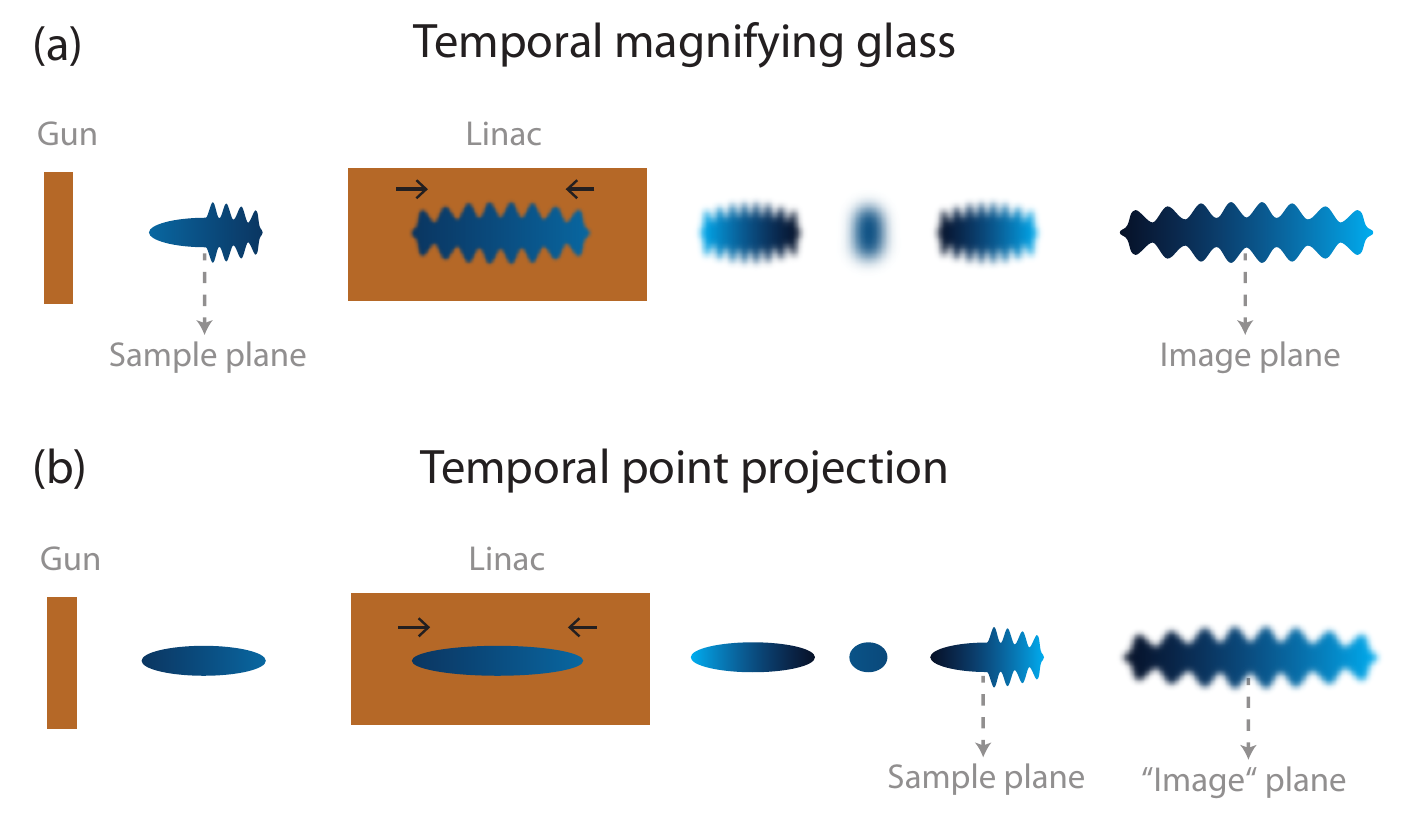}
\caption{Temporal imaging: (a) magnifying glass mode and (b) point projection mode. Both cases make use of a linac as a longitudinal lens in order to magnify temporal structures in a beam.}
\label{fig:cartoon_both}
\end{figure}

Here we describe a method for achieving a high ($>$ 10x) temporal magnification between the sample and a deflector using an accelerating cavity (linac) as a longitudinal lens and therefore proportionally improving the resolution of streak-mode time-resolved electron scattering. We use a simple model of the beam dynamics in an RF cavity to estimate the longitudinal lens parameters and then discuss two designs for obtaining temporal magnification. In the first, a linac is used as a magnifying glass to image the temporal profile of the beam at the sample plane to the principal plane of the deflector (Fig.\,\ref{fig:cartoon_both}a); while in the second scheme, the sample is placed shortly after a temporal focus where a strongly correlated longitudinal phase space distribution ($t-\gamma$ chirp) allows a shadow of the temporal dynamics to be projected downstream (Fig.\,\ref{fig:cartoon_both}b). As we will see, the later technique does not produce a proper image but it greatly simplifies the experimental setup and reduces the impact of space charge. Finally, we validate our concept using start-to-end particle-tracking simulations of a realistic MeV electron beam-line based on the Pegasus facility at UCLA\cite{maxson_direct_2017}. The study of this practical example is useful to establish the limits of the technique and highlight the relative merits of the two modalities.

\section{Theory}
It is instructive to start from a review on how a linear accelerator cavity can be used as a temporal lens for a bunched electron beam. In the discussion of this paper we consider radiofrequency based cavities, but the formalism could be extended to any accelerating element in an electron beamline (i.e. including THz or laser-based). For simplicity, we also restrict ourselves to a thin-lens model and leave the more complicated cases to particle tracking simulations.

Linear transport through TM010 cavities has been derived in the literature based on the momentum transfer to the beam from the RF wave \cite{pasmans_microwave_2013}. Neglecting transverse effects and assuming that the duration of the beam is short compared to the period of the resonant mode in the cavity, we can write for the change in momentum $\Delta p$ of a particle going through an RF cavity
\begin{equation}
\Delta p_z = \frac{e V_0}{\beta c}\left(\sin \phi_0 + \omega \frac{\zeta}{\beta c}\cos(\phi_0) \right)
\label{Eq:deltap}
\end{equation}
where $V_0$ is the accelerating voltage of the cavity and the bunch centroid passes through the center of the cavity at phase $\phi_0$. Throughout this paper we use the convention for which $\phi_0 = \pi/2$ is the phase for maximum acceleration. $\zeta$ is the longitudinal coordinate of the particle referred to the center of the beam. In Eq. \ref{Eq:deltap} the first term represents the beam change of mean energy and the second term gives the `restoring force' (i.e. $R_{65}$ in beam transport notation). In such a thin lens approximation the linear transport of the phase space vector $\langle \zeta, \Delta_{\gamma\beta} \rangle$ (with $\Delta_{\gamma\beta}=(p_z-p_{zc})/mc$ referenced to the center trajectory) is given by:
\begin{equation}
R_{RF} = \begin{pmatrix}
1 & 0 \\
\frac{e V_0 \omega/c}{m \beta^2 c^2}\cos \phi_0 & 1
\label{Eq:thinlens}
\end{pmatrix}
\end{equation}

After propagation downstream of an element represented by Eq.\,\ref{Eq:thinlens}, a collimated beam will form a focus a distance $f$ downstream of the lens, where
\begin{equation}
\frac{1}{f} = \frac{eV_0\omega}{m c^3 \gamma^3 \beta^4} \cos \phi_0
\end{equation}
and we have assumed that the linac is followed by a simple drift of length $L$. To give a numerical example, an ideal zero-length S-band ($\nu$ = 2.856\,GHz) 1\,MV cavity would have an effective focal length $f = 1.5$\,m for a 3\,MeV electron beam injected at zero-crossing phase (i.e. $\phi_0$ = 0).

\subsection{A magnifying glass}

The most direct analogy for a magnifying glass in longitudinal phase-space is an accelerating structure operating near its “zero-crossing”. Such structures are often used by the UED community to compress the beam and counteract the space-charge induced beam expansion \cite{van_oudheusden_compression_2010,gliserin_compression_2012,maxson_direct_2017}. Here we consider using one such cavity located after the sample at a distance such that a (magnified) temporal replica of the beam is recreated at the streaking plane. 

A matrix-based description of the transport through the magnifying glass can be written as:
\begin{equation}
\label{eq:Magnifying_Matrix}
R_{tot} =
\begin{pmatrix}
1 & \frac{L_2}{\beta_2^2\gamma_2^3} \\
0 & 1
\end{pmatrix}
R_{RF}
\begin{pmatrix}
1 & \frac{L_1}{\beta_1^2\gamma_1^3} \\
0 & 1
\end{pmatrix}
\end{equation}
where we have allowed the energy to change inside the RF cavity ($\phi_0\neq 0$). Choosing the cavity voltage to satisfy the imaging condition, we find that the magnification is: $M=\left(L_2/L_1\right) \left(\beta_1^2\gamma_1^3 / \beta_2^2 \gamma_2^3\right)$ where $L_1$ and $L_2$ are drift distances to and from the cavity.  


To achieve high magnification with a single lens one would typically maximize ($L_2/L_1$), but for realistic cavity voltages the total length $L_1+L_2$ can become quite long. This is because for a relativistic beam the dispersion of free space is small and the effective focal distance of the lens is long. 

Thus, an alternative to alleviate space constraints is to operate slightly off-crest so that the beam looses energy, adding dispersion, and increasing magnification as $\left(\beta_1^2\gamma_1^3 / \beta_2^2 \gamma_2^3\right)$. One important assumption in this analysis is that Eq.\,\ref{Eq:thinlens} still applies (i.e. that the phase slippage in the cavity remains independent of $\phi_0$). Another option, significantly raising the level of complexity in terms of hardware, is to control the beamline dispersion by adding a magnetic chicane.


\subsection{Point-projection}

Instead of placing the sample plane before the lens, as with the magnifying glass, it is also possible to place the sample shortly after the temporal focus where the beam is strongly correlated in longitudinal phase space so that a `shadow' of the sample is projected downstream as the beam expands (Fig.\,\ref{fig:cartoon_both}(b)). In addition to providing magnification between the sample and deflector planes, this scheme has the benefit of reducing the peak current which needs to be drawn from the cathode since the bunch is compressed near the sample plane (with the drawback of a smaller temporal observation window). Furthermore the strong $t-\gamma$ correlation means that the final image could be formed by using a spectrometer instead of a deflecting cavity. This method necessarily produces a defocused image, and so the limiting resolution ($\sigma_\text{psf}$) strongly depends on the beam quality.

In order to derive the magnification and resolution of the point-projection scheme we consider a beam after it has been focused by the accelerating structure and is drifting to the deflector plane. In particular, if we write the longitudinal phase space density $f_\text{s}(\zeta,\Delta_{\gamma \beta})$ at the sample plane, we can evaluate the resolution of a point-like sample with transmission contrast $c$ centered around position $\zeta_s$ after transport for a distance $L$ (neglecting space-charge):
\begin{equation}
g(\zeta; L) =\int_{-\infty}^{\infty} d\Delta_{\gamma \beta} f_s(\zeta-L\frac{\Delta_{\gamma \beta}}{\beta^2\gamma^3},\Delta_{\gamma \beta}) c(\zeta-L \Delta_{\gamma \beta}-\zeta_\text{s})
\end{equation}
Assuming an initial gaussian profile for the distribution function $f$ and a delta-function like behavior for $c$, by calculating the moments of $g$ we can estimate the magnification and resolution of the system. Expressing the results in terms of the beam moments at the sample gives:
\begin{align}
	M&=\frac{\langle \zeta \rangle}{\zeta_s}=1+\frac{L}{\beta^2\gamma^3}\frac{\langle \zeta \Delta_{\gamma \beta} \rangle_\text{s}}{\langle \zeta^2 \rangle_\text{s}} \\
    \sigma_\text{psf}&=\frac{\sqrt{\langle \zeta^2 \rangle-\langle \zeta \rangle^2}}{M}=\frac{L}{\beta^2\gamma^3} \sqrt{\langle \zeta^2 \rangle_\text{s}}\frac{1}{M}
    \label{Eq:GaussianMagRes}
\end{align}
The resolution is perfect only for $L=0$ ( i.e. we are directly looking at the sample) and rapidly gets worse until the dispersion is balanced by the increased magnification.

In the far-field the resolution can be expressed in terms of the longitudinal emittance and the RMS pulse-length at the temporal focus ($z_f$):
\begin{equation}
\lim_{\text{M}>>1}{\sigma_\text{psf}} \rightarrow \sigma_{\zeta}|_{z=z_f}\sqrt{1+\frac{\beta^2\gamma^3}{L}\frac{\langle \zeta^2 \rangle_f}{\epsilon}+...}
\label{Eq:GaussianRes}
\end{equation}

\begin{figure}[h]
\centering
\includegraphics[width=\columnwidth]{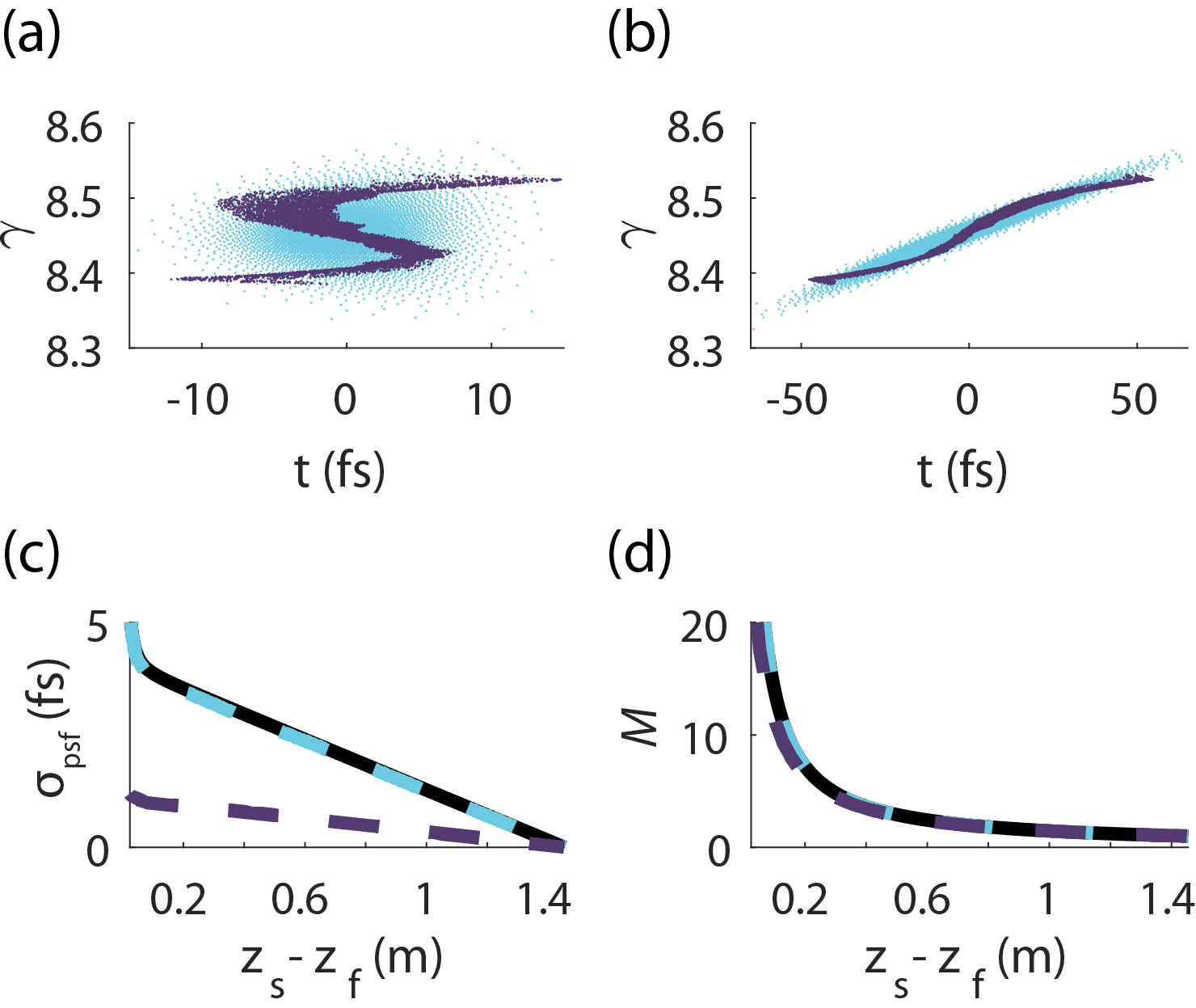}
\caption{Comparison of point-projection imaging between cases using a nonlinear phase space (purple) and cases using the equivalent Gaussian (blue). Longitudinal phase space of the two distributions at (a) the temporal focus ($z_f$) and (b) after a 0.1m drift ($z_f+0.1$). Placing a sample at a location ($z_s$) where the phase space is highly correlated allows imaging with an rms point spread function shown in (c) and the magnification shown in (d). The black line showing the calculations of Eq.\,\ref{Eq:GaussianMagRes} and Eq.\,\ref{Eq:GaussianRes} overlaps with the blue line for the Gaussian distribution.
}
\label{fig:point_proj_sample_location}
\end{figure}

Equation (\ref{Eq:GaussianRes}) only holds for Gaussian initial phase space distributions, but it is interesting to consider what happens for more complicated phase-spaces. Imagine that at the temporal focus an RF-induced higher order curvature folds the phase space\cite{zeitler_linearization_2015} such that $f$ follows a polynomial curve $\zeta|_{z=z_f} = a_2 \Delta_{\gamma \beta}^2+a_3 \Delta_{\gamma \beta}^3+...$ (Fig.\,\ref{fig:point_proj_sample_location}a). As the beam disperses the curve will get stretched like $\zeta|_{z=z_f+L}=\zeta|_{z=z_f}+\frac{L}{\beta^2\gamma^3} \Delta_{\gamma \beta}$ and for large enough $L$ this linear term will begin to dominate the distribution. If the sample is placed sufficiently far away (Fig.\,\ref{fig:point_proj_sample_location}b) then each time-slice will have much better rms resolution than $\sigma_{\zeta}|_f$; however, at the detector plane, the magnification of the slices at the head and tail of the distribution will be different.

In Fig.\,\ref{fig:point_proj_sample_location}(c,d) we compare resolution and magnification for a realistic beam and an equivalent Gaussian as function of the distance between the sample and the crossover focal plane. The black lines are the prediction of Eq.\ref{Eq:GaussianMagRes}, which matches well with the light-blue lines from a particle tracking simulation of a Gaussian beam having emittance $\epsilon$ and temporal focus $\sigma_{\zeta}|_{z=z_f}$. The purple line shows the same plot for a 100\,fC particle beam tracked through an RF linac and having the same $\epsilon$ and $\sigma_{\zeta}|_{z=z_f}$ as the Gaussian beams. Because the purple phase space is folded it has significantly less slice energy spread in \ref{fig:point_proj_sample_location}$b$ than the equivalent Gaussian. However the chirp of the beam also varies with $t$ so that point projection imaging using this beam will result in a non-linear mapping from $t_i \rightarrow t_f$ (causing a distortion, but not blurring, of the image). In any case, the general result holds that for large magnification the temporal resolution of the point-projection scheme is limited to roughly the pulse length at the crossover focus ($\sigma_{\zeta}|_{z=z_f}$).

\section{Results and discussion}

\begin{figure}
\centering
\includegraphics[width=\columnwidth]{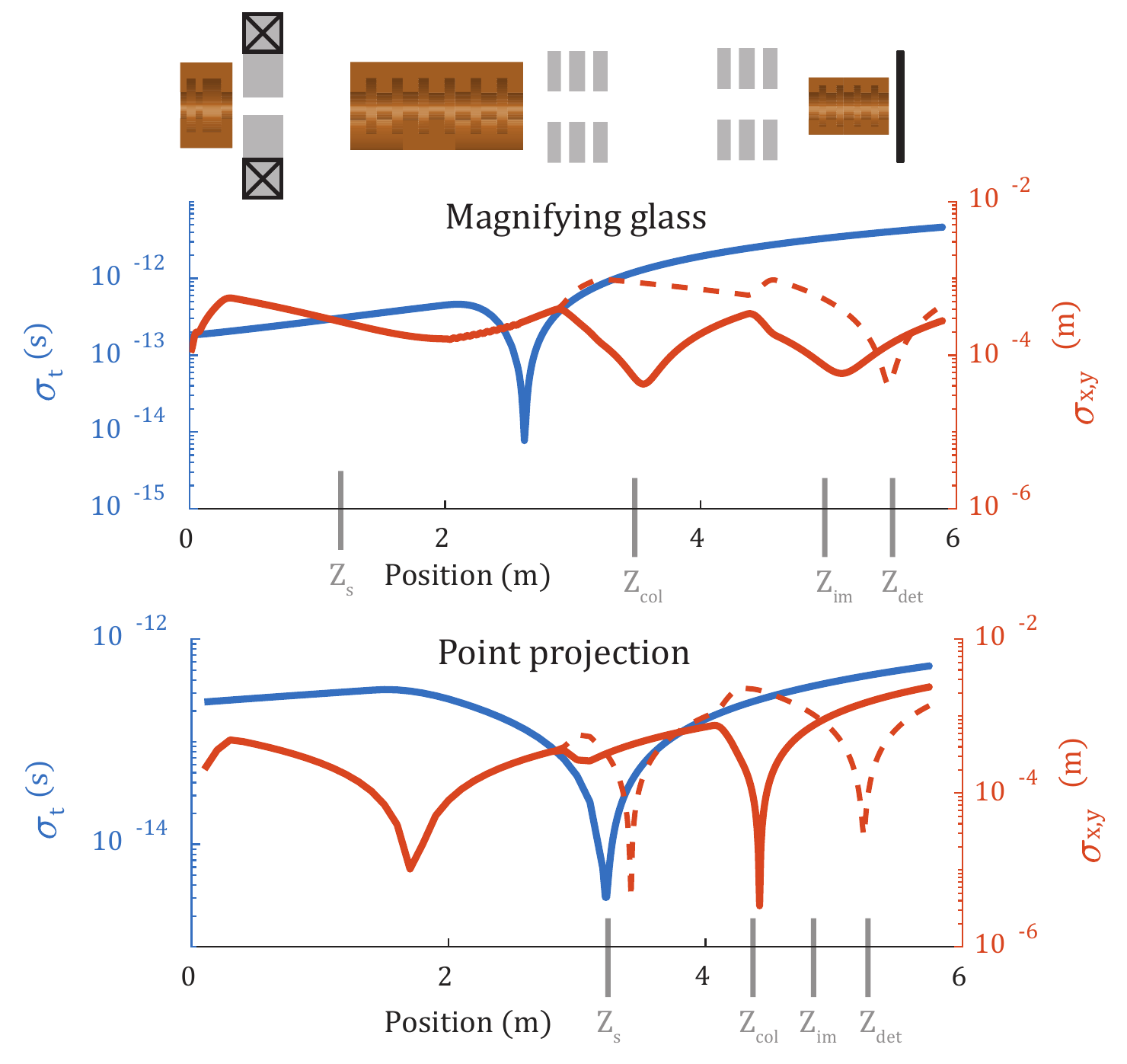}
\caption{Schematic of the beamline used to study the feasibility of temporal imaging. The blue line shows $\sigma_t$ and the orange lines show $\sigma_{x,y}$ (solid and dashed respectively). The beam envelopes are shown without any kick from the sample or deflecting cavity such that collimator and detector locations ($z_\text{col}$ and $z_\text{det}$ respectively) appear near waists.}
\label{fig:beamlines}
\end{figure}

To illustrate the feasibility of temporal imaging we design two new configurations of the Pegasus beamline at UCLA which can be used to obtain 10x temporal magnification. The start-to-end simulations include space-charge and use field-maps for the RF structures which have recently been validated with experimental measurements of pulse compression \cite{maxson_direct_2017}.  

\subsection{Image forming process}
A schematic of the two configurations is shown in Fig.\,\ref{fig:beamlines} with relevant lens parameters listed in Table\,\ref{tab:lens_parameters}. The beamline starts with a 1.6 cell RF photogun having a 70\,MV/m peak field which accelerates the electrons to $\gamma=7.1$. The gun is immediately followed by a solenoid which focuses into the linac for temporal imaging. After the linac are two sets of quadrupoles, a collimator, and a  200\,kV x-band deflector used for creating the streaked images.

\begin{table}
\centering
\begin{tabular} {|c |c| c| c|}
\hline
 &Mag glass & Point proj & Unit\\
\hline
$\phi_0$ & -8 & 7.6 &  deg \\
$V$ & 8.9 & 3.9 & MV \\
$z_\text{s}$ & 1.3 & 3.36 & m \\
$z_\text{linac}$ & 2.3 & 1.72 & m \\
$z_\text{col}$ & 3.21 & 4.695 & m \\
$z_\text{im}$ & 4.85 & 4.85 & m \\
\hline
\end{tabular}
\caption{Comparison of temporal lens parameters between the magnifying glass and point-projection configurations}
\label{tab:lens_parameters}
\end{table}

We form a streaked image in two stages: first the scattering from the sample is projected onto a collimator so that we can create contrast by choosing a slice of k-space; and secondly the beam is refocused so that the time-dependent kick from the x-band deflector can be seen on the detector. For both process we use a quadrupole triplet to help image an angular kick from the scattering/deflection plane to the collimator/detector plane. Separate from these processes, we use the linac to create temporal magnification between the sample and the deflector.

The resolution of the streaked images is limited by our ability to resolve small angular kicks from the sample/deflector. We can only distinguish scattering angles larger that the uncorrelated angular spread of the electron beam, which for an emittance $\epsilon$ and spot-size $\sigma_x$  (at the scattering/deflection plan) is $\theta\approx\epsilon/\sigma_x$. Thus, for fixed emittance there is a trade-off between the size of the beam (controlled by the quadrupole triplets) and the angular resolution; and, all other things equal, a better emittance is directly proportional to a better resolution.

Better emittance is most easily achieved by reducing the source size at the cathode\,\cite{maxson_direct_2017},  but this comes at the cost of increased charge density, and thus space-charge forces. The cathode emittance we can obtain is thus constrained by the need to generate a high-current electron beam in order to provide a sufficient number of scattered electrons per unit time to overcome shot-noise in the final streaked image. This can be appreciated by comparing the beam parameters for the magnifying glass and point-projection configurations  (Table\,\ref{tab:parameters}).



The primary difference between the configurations is that the magnifying-glass configuration locates $z_\text{obj}$ before the linac and so can produce a real image at $z_\text{im}$, while the point-projection configuration only casts a shadow of the object. The magnifying glass produces a more accurate temporal replica at $z_\text{im}$, but it doesn't compress the beam before the sample, and so it must draw more charge off the cathode in order to provide the same current. In order to accommodate the larger cathode current the magnifying-glass configuration has a larger source size and therefore a larger emittance. Note however, that the small emittance from the point-projection configuration is not preserved along the beamline due to chromatic aberration from the large energy spread applied by the linac.

\subsection{Temporal resolution}
\begin{figure}
\centering
\includegraphics[width=\columnwidth]{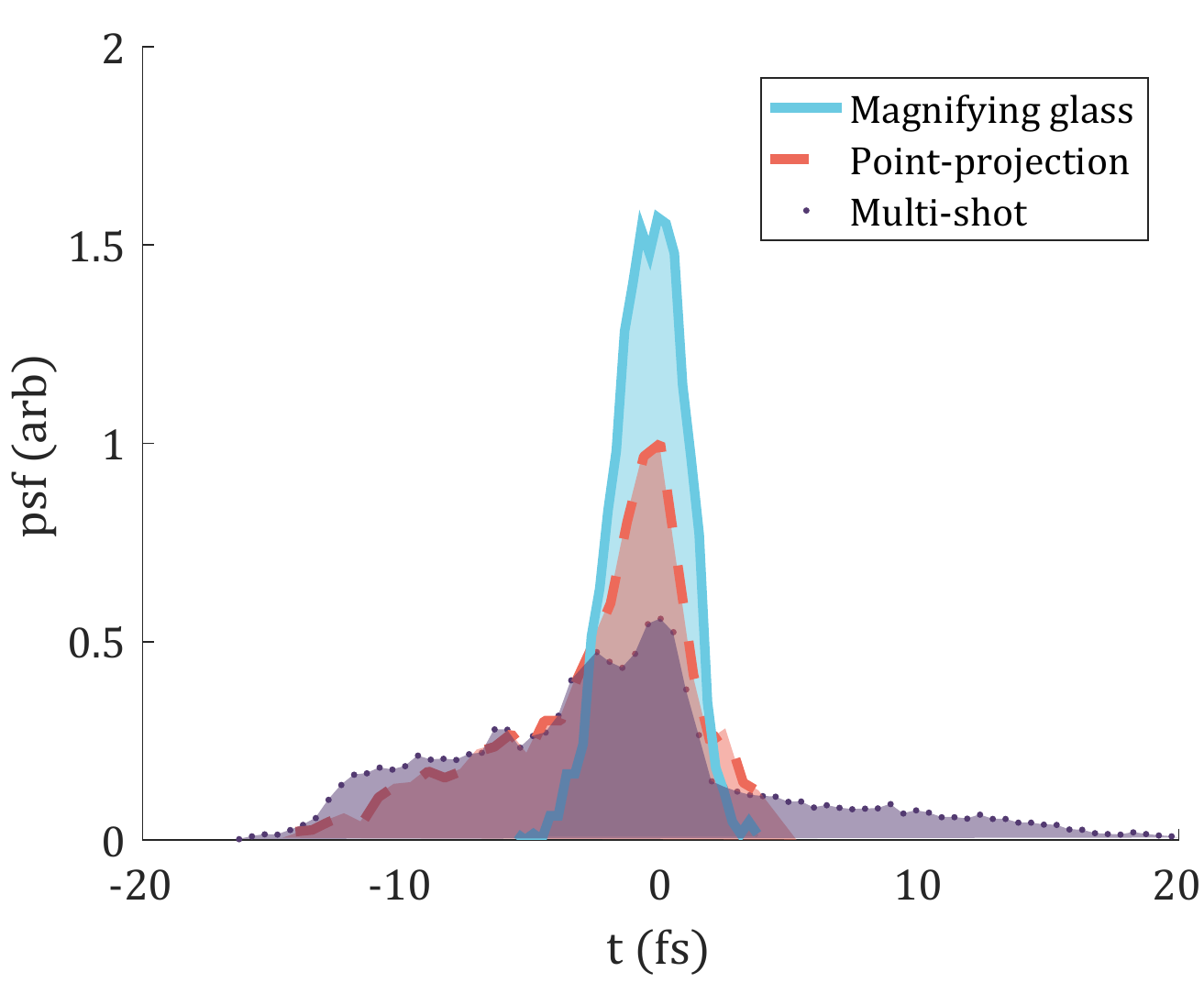}
\caption{Simulated point spread function for several configurations. The magnifying-glass and point-projection curves show the resolution after the deflector, while multi-shot shows the beam's temporal distribution at full compression (which is the resolution for conventional un-streaked UED).}
\label{fig:psfs}
\end{figure}

The result of particle tracking through the two configurations (including the deflector) can be summarized by their point-spread functions as shown in Fig.\,\ref{fig:psfs}. They can be compared to an un-streaked, multi-shot approach in which the temporal resolution is ultimately limited by the pulse width at full compression. The multi-shot and point-projection techniques have similar blurring, as expected from Fig.\,\ref{fig:point_proj_sample_location}, because the phase-space correlation is primarily determined by the RF curvature. The multi-shot approach, however, benefits from using a fully compressed bunch and thus has nearly 50x more charge per time-slice at the cost of increased slice energy spread and the practical limits set by shot-to-shot timing jitter. The magnifying glass configuration has by far the best temporal resolution (1.4\,fs rms) and is only limited by nonlinear space-charge interactions. 

\begin{table}
\begin{tabular} {|c |c|c| c| c|}
\hline
 &Z& Mag glass & Point proj & Unit\\
\hline
$\sigma_{x,y}$& $z_0$ & 75 & 20 & $\mu$m \\
Charge& & 1000 & 100 & fC \\
$I$& $z_\text{s}$ & 1 & 1 & A \\
$\sigma_{t}$& $z_\text{s}$& 350 & 35 & fs \\
$\sigma_{x,y}$&$z_\text{s}$ & 100 & 100 & $\mu$m \\
$\epsilon_{x}$&$z_\text{s}$ & 70 & 60 & nm \\
$M$& $z_\text{im}$ & 10 & 10 &  \\
$\sigma_\text{psf}$&$z_\text{det}$ & 1.4 & 2.5 & fs \\
\hline
\end{tabular}
\caption{Comparison of beam parameters between the magnifying glass and point-projection configurations}
\label{tab:parameters}
\end{table}

In the absence of space-charge the magnifying glass can achieve attosecond resolution, but because the beam has to be imaged through a temporal focus the space charge causes significant blurring. By comparison, the point-projection scheme is relatively indifferent to space charge: firstly, because obtaining the same current at the sample requires lower peak current; and secondly because the sample is located after the crossover where the space charge force is mostly linear.

\begin{figure}[t]
\centering
\includegraphics[width=\columnwidth]{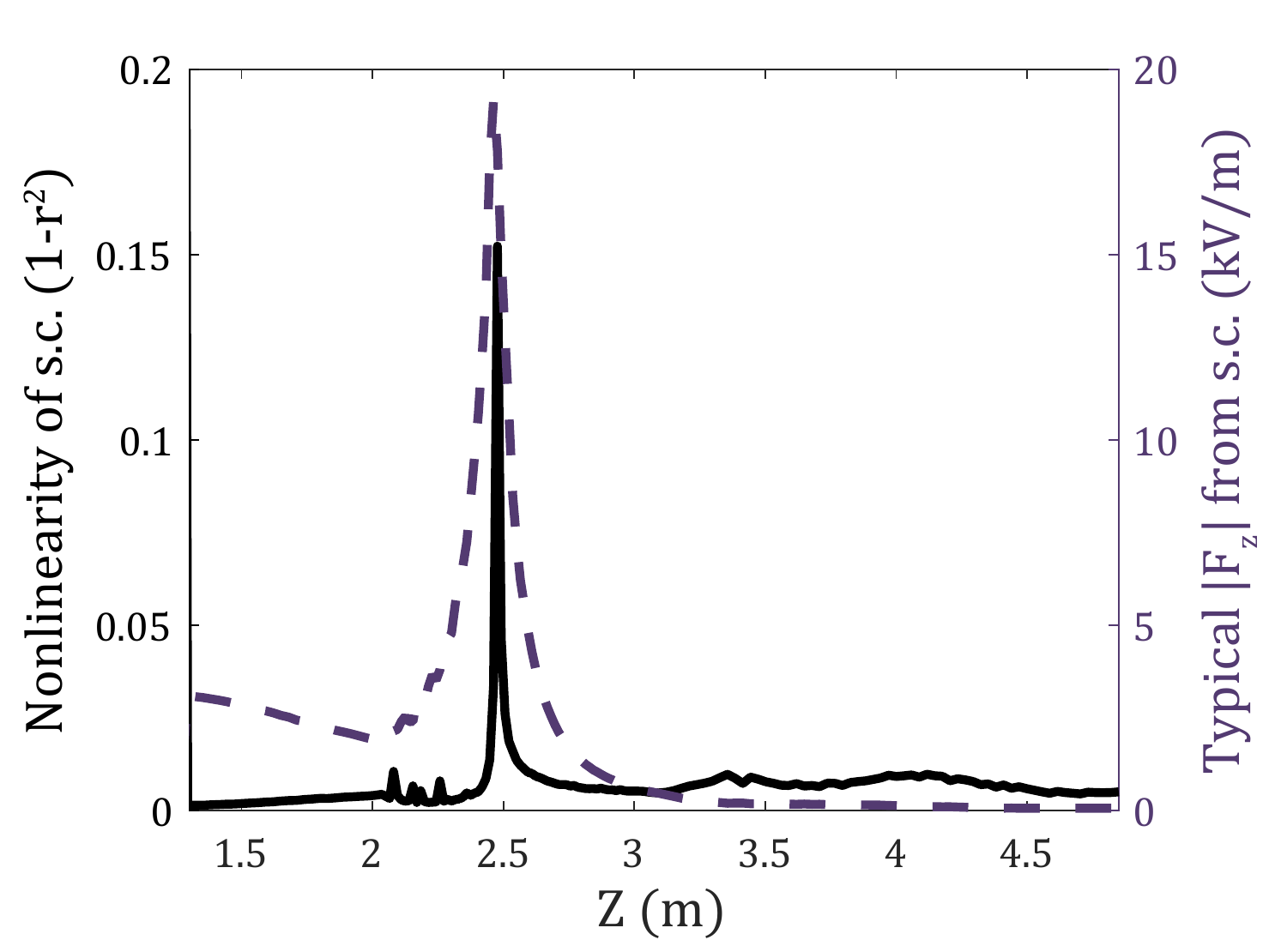}
\caption{Nonlinear space charges forces from the sample plane to the image. The left axis (solid line) uses the Pearson $r$ coefficent to indicate linearity of the longitudinal space charge force. The right axis (dashed line) shows the median amplitude of the space charge force, which spikes at full compression.}
\label{fig:nonlinearity}
\end{figure}

The problem of nonlinear space charge repulsion blurring the beam can be seen  in Fig.\,\ref{fig:nonlinearity} where we represent nonlinearity via the Pearson $r$ coefficient of a regression on $z-E_z$ for particles which started within $100$\,fs of the beam centroid at the sample plane. Near the sample the space-charge force is linear, as expected for the blowout regime\cite{luiten_how_2004}, and largely benign; but as the beam approaches the temporal crossover the beam current profile is no longer uniform (Fig.\,\ref{fig:point_proj_sample_location}(a)) and the force is simultaneously strong and nonlinear. In this example space charge decreases the rms resolution from $<100$\,as to $1.4$\,fs. In principle, this is not a fundamental limit: for example, if the temporal crossover occurs far enough after the linac that the beam can be defocused then the magnitude of the space charge kick can be reduced. 

\subsection{Simulated streaking patterns}
Ultrafast streaking such as described in this paper could be used to observe near-field phenomena driven by an intense laser pulse  \cite{schiffrin_optical-field-induced_2013,cesar_nonlinear_2017,morimoto_diffraction_2018}, for which it is necessary to detect small changes in a periodic scattering signal. To understand how we can see such a signal we consider a simple scattering model in which the sample provides a sinusoidal kick $A \sin(\omega t)$. The goal of the beamline design is to minimize the kick magnitude, $A$, required to generate contrast while retaining sufficient temporal resolution to resolve the frequency $\omega$. 

Using this scattering model and the collimators indicated in Fig.\,\ref{fig:beamlines} we can simulate the entire image forming process for an $A=0.5$\,mrad, $f=1.25\cdot10^{14}$\,hz signal (corresponding to a 2.4$\mu$m laser). The electron beams we use in our simulation (see Table \ref{tab:parameters}) have about 50,000 electrons per optical cycle; however we expect that only a small fraction of these will be scattered. Since the scattered fraction is application dependent, we do not simulate it here. Instead we simulate a small number of macro-particles sampled from a low-discrepancy data set. Thus, these results are suitable to judge the temporal and angular resolution of the image, but the effects shot-noise have to be considered on a case-by-case basis. We expect that these results are directly comparable to a single-shot diffraction pattern with 4\,fC per shot.

\begin{figure}
\centering
\includegraphics[width=\columnwidth]{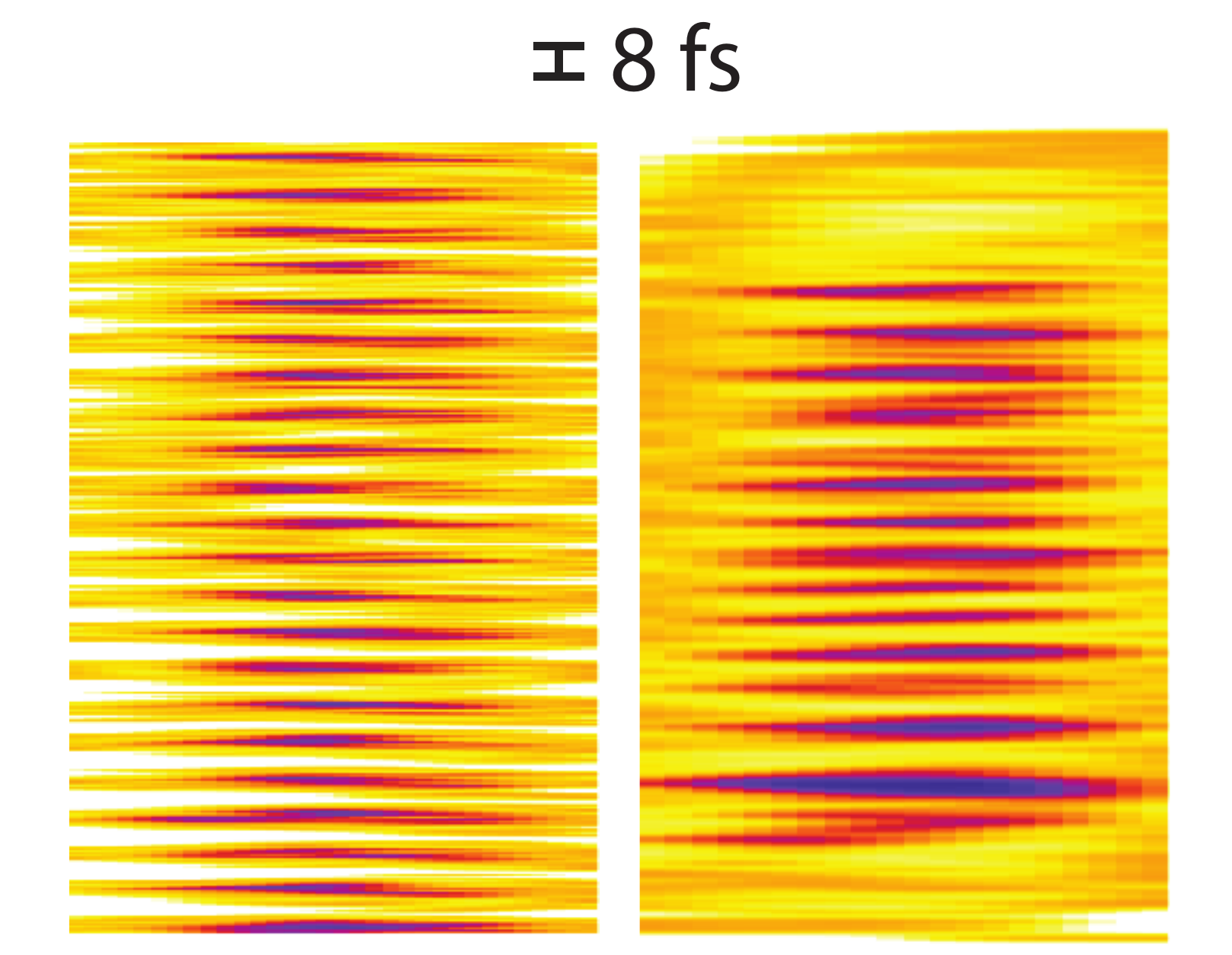}
\caption{Simulated images of streaked electrons after a $0.5$\,mrad modulation from a 2.4um laser. The magnifying glass (left) performs better than the point-projection configuration (right). In both cases the temporal magnification is 10 so that after the deflector 8\,fs corresponds to 140\,$\mu$m.}
\label{fig:beam_images}
\end{figure}

In a side-by-side comparison we can see that both methods are able to resolve the high-frequency signal (Fig.\,\ref{fig:beam_images}), however we can clearly see advantages of the magnifying glass: better resolution, larger field-of-view, and less distortion of the image. This is because the point-projection technique does not form a real image and so the temporal-resolution is distorted by higher-order RF curvature (see Fig.\,\ref{fig:point_proj_sample_location}) and the field-of-view is limited. Nonetheless, the point-projection configuration remains appealing for its simplicity, especially when considering that the $t-\gamma$ correlation means the deflecting cavity could be replaced by a high resolution energy spectrometer \cite{baum_femtosecond_2008} without loosing any resolution.

\section{Conclusions}

We have studied a new concept based on the use of an RF cavity as a longitudinal lens to provide 10x temporal magnification and increase the resolution of a single-shot electron streak camera. Starting from simple calculations and progressing to full simulations we have illustrated the physical mechanisms which influence and ultimately limit the temporal imaging technique. Our results suggest an (rms) resolution of 1.4\,fs and field-of-view of $1$\,ps can be achieved in the magnifying glass configuration. The alternative point-projection scheme is more limited, but has an attractive simplicity: the deflecting cavity can be replaced by a spectrometer.



Both streaking techniques rely on a data acquisition modality which is well-suited to studying strongly scattering ultrafast phenomena. The streaking methods (as opposed to a bi-dimensional diffraction pattern) typically constrain us to acquire diffraction patterns along a single scattering vector, and, depending on the modality, with a reduced signal-to-noise ratio. Nevertheless, these methods do allow us to extend our time-resolution to the optical-scale. This is advantageous both for studying diffraction patterns and for studying laser-electron interactions in below-ionization phenomena \cite{cesar_nonlinear_2017, morimoto_diffraction_2018,barwick_photon-induced_2009}, similar to how streaked photoelectrons are used to study photoionization \cite{blaga_imaging_2012} or electrically induced conductivity \cite{schiffrin_optical-field-induced_2013}. We envision streaking electrons over a $1$\,ps field-of-view with sub-optical resolution in order to bridge the gap between attoscience and conventional time-resolved imaging techniques. 

\section{Acknowledgements}
This work has been partially supported by the National Science Foundation under Grant No. DMR-1548924.


\bibliography{bib}

\begin{thebibliography}{10}
\expandafter\ifx\csname url\endcsname\relax
  \def\url#1{\texttt{#1}}\fi
\expandafter\ifx\csname urlprefix\endcsname\relax\def\urlprefix{URL }\fi
\expandafter\ifx\csname href\endcsname\relax
  \def\href#1#2{#2} \def\path#1{#1}\fi

\bibitem{zewail_4d_2006}
A.~H. Zewail,
  \href{https://www.annualreviews.org/doi/10.1146/annurev.physchem.57.032905.104748}{4d
  ultrafast electron diffraction, crystallography, and microscopy}, Annu. Rev.
  Phys. Chem. 57~(1) (2006) 65--103.
\newblock \href {https://doi.org/10.1146/annurev.physchem.57.032905.104748}
  {\path{doi:10.1146/annurev.physchem.57.032905.104748}}.
\newline\urlprefix\url{https://www.annualreviews.org/doi/10.1146/annurev.physchem.57.032905.104748}

\bibitem{dwyer_femtosecond_2006}
J.~R. Dwyer, C.~T. Hebeisen, R.~Ernstorfer, M.~Harb, V.~B. Deyirmenjian, R.~E.
  Jordan, R.~J.~D. Miller,
  \href{http://rsta.royalsocietypublishing.org/content/364/1840/741}{Femtosecond
  electron diffraction: ‘making the molecular movie’}, Philosophical
  Transactions of the Royal Society of London A: Mathematical, Physical and
  Engineering Sciences 364~(1840) (2006) 741--778.
\newblock \href {https://doi.org/10.1098/rsta.2005.1735}
  {\path{doi:10.1098/rsta.2005.1735}}.
\newline\urlprefix\url{http://rsta.royalsocietypublishing.org/content/364/1840/741}

\bibitem{king_ultrafast_2005}
W.~E. King, G.~H. Campbell, A.~Frank, B.~Reed, J.~F. Schmerge, B.~J. Siwick,
  B.~C. Stuart, P.~M. Weber,
  \href{https://aip.scitation.org/doi/abs/10.1063/1.1927699}{Ultrafast electron
  microscopy in materials science, biology, and chemistry}, Journal of Applied
  Physics 97~(11) (2005) 111101.
\newblock \href {https://doi.org/10.1063/1.1927699}
  {\path{doi:10.1063/1.1927699}}.
\newline\urlprefix\url{https://aip.scitation.org/doi/abs/10.1063/1.1927699}

\bibitem{sciaini_femtosecond_2011}
G.~Sciaini, R.~J.~D. Miller,
  \href{http://stacks.iop.org/0034-4885/74/i=9/a=096101}{Femtosecond electron
  diffraction: heralding the era of atomically resolved dynamics}, Rep. Prog.
  Phys. 74~(9) (2011) 096101.
\newblock \href {https://doi.org/10.1088/0034-4885/74/9/096101}
  {\path{doi:10.1088/0034-4885/74/9/096101}}.
\newline\urlprefix\url{http://stacks.iop.org/0034-4885/74/i=9/a=096101}

\bibitem{hall_future_2014}
E.~Hall, S.~Stemmer, H.~Zheng, Y.~Zhu, G.~Maracas,
  \href{https://www.osti.gov/biblio/1287380}{Future of {Electron} {Scattering}
  and {Diffraction}}, Tech. rep., US Department of Energy, Washington, DC
  (United States) (Feb. 2014).
\newblock \href {https://doi.org/10.2172/1287380} {\path{doi:10.2172/1287380}}.
\newline\urlprefix\url{https://www.osti.gov/biblio/1287380}

\bibitem{weathersby_mega-electron-volt_2015}
S.~P. Weathersby, G.~Brown, M.~Centurion, T.~F. Chase, R.~Coffee, J.~Corbett,
  J.~P. Eichner, J.~C. Frisch, A.~R. Fry, M.~Gühr, N.~Hartmann, C.~Hast,
  R.~Hettel, R.~K. Jobe, E.~N. Jongewaard, J.~R. Lewandowski, R.~K. Li, A.~M.
  Lindenberg, I.~Makasyuk, J.~E. May, D.~McCormick, M.~N. Nguyen, A.~H. Reid,
  X.~Shen, K.~Sokolowski-Tinten, T.~Vecchione, S.~L. Vetter, J.~Wu, J.~Yang,
  H.~A. Dürr, X.~J. Wang,
  \href{https://aip.scitation.org/doi/abs/10.1063/1.4926994}{Mega-electron-volt
  ultrafast electron diffraction at {SLAC} {National} {Accelerator}
  {Laboratory}}, Review of Scientific Instruments 86~(7) (2015) 073702.
\newblock \href {https://doi.org/10.1063/1.4926994}
  {\path{doi:10.1063/1.4926994}}.
\newline\urlprefix\url{https://aip.scitation.org/doi/abs/10.1063/1.4926994}

\bibitem{van_oudheusden_compression_2010}
T.~van Oudheusden, P.~L. E.~M. Pasmans, S.~B. van~der Geer, M.~J. de~Loos,
  M.~J. van~der Wiel, O.~J. Luiten,
  \href{https://link.aps.org/doi/10.1103/PhysRevLett.105.264801}{Compression of
  {Subrelativistic} {Space}-{Charge}-{Dominated} {Electron} {Bunches} for
  {Single}-{Shot} {Femtosecond} {Electron} {Diffraction}}, Phys. Rev. Lett.
  105~(26) (2010) 264801.
\newblock \href {https://doi.org/10.1103/PhysRevLett.105.264801}
  {\path{doi:10.1103/PhysRevLett.105.264801}}.
\newline\urlprefix\url{https://link.aps.org/doi/10.1103/PhysRevLett.105.264801}

\bibitem{musumeci_laser-induced_2010}
P.~Musumeci, J.~T. Moody, C.~M. Scoby, M.~S. Gutierrez, M.~Westfall,
  \href{http://scitation.aip.org/content/aip/journal/apl/97/6/10.1063/1.3478005}{Laser-induced
  melting of a single crystal gold sample by time-resolved ultrafast
  relativistic electron diffraction}, Applied Physics Letters 97~(6) (2010)
  063502.
\newblock \href {https://doi.org/10.1063/1.3478005}
  {\path{doi:10.1063/1.3478005}}.
\newline\urlprefix\url{http://scitation.aip.org/content/aip/journal/apl/97/6/10.1063/1.3478005}

\bibitem{Mourou&Williamson}
G.~Mourou, S.~Williamson, \href{https://doi.org/10.1063/1.93316}{Picosecond
  electron diffraction}, Applied Physics Letters 41~(1) (1982) 44--45.
\newblock \href {http://arxiv.org/abs/https://doi.org/10.1063/1.93316}
  {\path{arXiv:https://doi.org/10.1063/1.93316}}, \href
  {https://doi.org/10.1063/1.93316} {\path{doi:10.1063/1.93316}}.
\newline\urlprefix\url{https://doi.org/10.1063/1.93316}

\bibitem{li:continuousUED}
R.~Li, W.~Huang, Y.~Du, L.~Yan, Q.~Du, J.~Shi, J.~Hua, H.~Chen, T.~Du, H.~Xu,
  et~al., Note: Single-shot continuously time-resolved mev ultrafast electron
  diffraction, Review of Scientific Instruments 81~(3) (2010) 036110.

\bibitem{musumeci_capturing_2010}
P.~Musumeci, J.~T. Moody, C.~M. Scoby, M.~S. Gutierrez, M.~Westfall, R.~K. Li,
  \href{https://aip.scitation.org/doi/abs/10.1063/1.3520283}{Capturing
  ultrafast structural evolutions with a single pulse of {MeV} electrons:
  {Radio} frequency streak camera based electron diffraction}, Journal of
  Applied Physics 108~(11) (2010) 114513.
\newblock \href {https://doi.org/10.1063/1.3520283}
  {\path{doi:10.1063/1.3520283}}.
\newline\urlprefix\url{https://aip.scitation.org/doi/abs/10.1063/1.3520283}

\bibitem{musumeci_double-shot_2017}
P.~Musumeci, D.~Cesar, J.~Maxson,
  \href{https://www.ncbi.nlm.nih.gov/pmc/articles/PMC5438282/}{Double-shot
  {MeV} electron diffraction and microscopy}, Struct Dyn 4~(4).
\newblock \href {https://doi.org/10.1063/1.4983390}
  {\path{doi:10.1063/1.4983390}}.
\newline\urlprefix\url{https://www.ncbi.nlm.nih.gov/pmc/articles/PMC5438282/}

\bibitem{qi_compressed_2018}
D.~Qi, C.~Yang, F.~Cao, J.~Liang, Y.~He, Y.~Yang, T.~Jia, Z.~Sun, S.~Zhang,
  \href{https://link.aps.org/doi/10.1103/PhysRevApplied.10.054061}{Compressed
  {Ultrafast} {Electron} {Diffraction} {Imaging} {Through} {Electronic}
  {Encoding}}, Phys. Rev. Applied 10~(5) (2018) 054061.
\newblock \href {https://doi.org/10.1103/PhysRevApplied.10.054061}
  {\path{doi:10.1103/PhysRevApplied.10.054061}}.
\newline\urlprefix\url{https://link.aps.org/doi/10.1103/PhysRevApplied.10.054061}

\bibitem{scoby_single-shot_2013}
C.~M. Scoby, R.~K. Li, E.~Threlkeld, H.~To, P.~Musumeci,
  \href{https://aip.scitation.org/doi/abs/10.1063/1.4776686}{Single-shot 35 fs
  temporal resolution electron shadowgraphy}, Appl. Phys. Lett. 102~(2) (2013)
  023506.
\newblock \href {https://doi.org/10.1063/1.4776686}
  {\path{doi:10.1063/1.4776686}}.
\newline\urlprefix\url{https://aip.scitation.org/doi/abs/10.1063/1.4776686}

\bibitem{akre_transverse_2001}
R.~Akre, L.~Bentson, P.~Emma, P.~Krejcik, A transverse rf deflecting structure
  for bunch length and phase space diagnostics, in: {PACS}2001. {Proceedings}
  of the 2001 {Particle} {Accelerator} {Conference} ({Cat}. {No}.01CH37268),
  Vol.~3, 2001, pp. 2353--2355 vol.3.
\newblock \href {https://doi.org/10.1109/PAC.2001.987379}
  {\path{doi:10.1109/PAC.2001.987379}}.

\bibitem{xiang_longitudinal--transverse_2010}
D.~Xiang, Y.~Ding,
  \href{https://link.aps.org/doi/10.1103/PhysRevSTAB.13.094001}{Longitudinal-to-transverse
  mapping for femtosecond electron bunch length measurement}, Phys. Rev. ST
  Accel. Beams 13~(9) (2010) 094001.
\newblock \href {https://doi.org/10.1103/PhysRevSTAB.13.094001}
  {\path{doi:10.1103/PhysRevSTAB.13.094001}}.
\newline\urlprefix\url{https://link.aps.org/doi/10.1103/PhysRevSTAB.13.094001}

\bibitem{fabianska_split_2014}
J.~Fabiańska, G.~Kassier, T.~Feurer,
  \href{https://www.nature.com/articles/srep05645}{Split ring resonator based
  {THz}-driven electron streak camera featuring femtosecond resolution},
  Scientific Reports 4 (2014) 5645.
\newblock \href {https://doi.org/10.1038/srep05645}
  {\path{doi:10.1038/srep05645}}.
\newline\urlprefix\url{https://www.nature.com/articles/srep05645}

\bibitem{zhao_terahertz_2018}
L.~Zhao, Z.~Wang, C.~Lu, R.~Wang, C.~Hu, P.~Wang, J.~Qi, T.~Jiang, S.~Liu,
  Y.~Shi, W.~Song, X.~Zhu, J.~Shi, Y.~Wang, L.~Yan, L.~Zhu, D.~Xiang, J.~Zhang,
  \href{http://arxiv.org/abs/1805.03923}{Terahertz streaking of few-femtosecond
  relativistic electron beams}, arXiv:1805.03923 [physics]ArXiv: 1805.03923.
\newline\urlprefix\url{http://arxiv.org/abs/1805.03923}

\bibitem{li_terahertz-based_2018}
R.~K. Li, M.~C. Hoffmann, E.~A. Nanni, S.~H. Glenzer, A.~M. Lindenberg, B.~K.
  Ofori-Okai, A.~H. Reid, X.~Shen, S.~P. Weathersby, J.~Yang, M.~Zajac, X.~J.
  Wang, \href{http://arxiv.org/abs/1805.01979}{Terahertz-based attosecond
  metrology of relativistic electron beams}, arXiv:1805.01979 [physics]ArXiv:
  1805.01979.
\newline\urlprefix\url{http://arxiv.org/abs/1805.01979}

\bibitem{chatelain_ultrafast_2012}
R.~P. Chatelain, V.~R. Morrison, C.~Godbout, B.~J. Siwick,
  \href{https://aip.scitation.org/doi/abs/10.1063/1.4747155}{Ultrafast electron
  diffraction with radio-frequency compressed electron pulses}, Appl. Phys.
  Lett. 101~(8) (2012) 081901.
\newblock \href {https://doi.org/10.1063/1.4747155}
  {\path{doi:10.1063/1.4747155}}.
\newline\urlprefix\url{https://aip.scitation.org/doi/abs/10.1063/1.4747155}

\bibitem{gao_full_2012}
M.~Gao, H.~Jean-Ruel, R.~R. Cooney, J.~Stampe, M.~d. Jong, M.~Harb, G.~Sciaini,
  G.~Moriena, R.~J.~D. Miller,
  \href{https://www.osapublishing.org/oe/abstract.cfm?uri=oe-20-11-12048}{Full
  characterization of {RF} compressed femtosecond electron pulses using
  ponderomotive scattering}, Opt. Express, OE 20~(11) (2012) 12048--12058.
\newblock \href {https://doi.org/10.1364/OE.20.012048}
  {\path{doi:10.1364/OE.20.012048}}.
\newline\urlprefix\url{https://www.osapublishing.org/oe/abstract.cfm?uri=oe-20-11-12048}

\bibitem{maxson_direct_2017}
J.~Maxson, D.~Cesar, G.~Calmasini, A.~Ody, P.~Musumeci, D.~Alesini,
  \href{https://link.aps.org/doi/10.1103/PhysRevLett.118.154802}{Direct
  {Measurement} of {Sub}-10 fs {Relativistic} {Electron} {Beams} with
  {Ultralow} {Emittance}}, Phys. Rev. Lett. 118~(15) (2017) 154802.
\newblock \href {https://doi.org/10.1103/PhysRevLett.118.154802}
  {\path{doi:10.1103/PhysRevLett.118.154802}}.
\newline\urlprefix\url{https://link.aps.org/doi/10.1103/PhysRevLett.118.154802}

\bibitem{franssen_improving_2017}
J.~G.~H. Franssen, O.~J. Luiten,
  \href{https://aca.scitation.org/doi/full/10.1063/1.4984104}{Improving
  temporal resolution of ultrafast electron diffraction by eliminating arrival
  time jitter induced by radiofrequency bunch compression cavities}, Structural
  Dynamics 4~(4) (2017) 044026.
\newblock \href {https://doi.org/10.1063/1.4984104}
  {\path{doi:10.1063/1.4984104}}.
\newline\urlprefix\url{https://aca.scitation.org/doi/full/10.1063/1.4984104}

\bibitem{pasmans_microwave_2013}
P.~L. E.~M. Pasmans, G.~B. van~den Ham, S.~F.~P. Dal~Conte, S.~B. van~der Geer,
  O.~J. Luiten,
  \href{http://www.sciencedirect.com/science/article/pii/S0304399112001817}{Microwave
  {TM}010 cavities as versatile 4d electron optical elements}, Ultramicroscopy
  127 (2013) 19--24.
\newblock \href {https://doi.org/10.1016/j.ultramic.2012.07.011}
  {\path{doi:10.1016/j.ultramic.2012.07.011}}.
\newline\urlprefix\url{http://www.sciencedirect.com/science/article/pii/S0304399112001817}

\bibitem{gliserin_compression_2012}
A.~Gliserin, A.~Apolonski, F.~Krausz, P.~Baum,
  \href{http://stacks.iop.org/1367-2630/14/i=7/a=073055}{Compression of
  single-electron pulses with a microwave cavity}, New J. Phys. 14~(7) (2012)
  073055.
\newblock \href {https://doi.org/10.1088/1367-2630/14/7/073055}
  {\path{doi:10.1088/1367-2630/14/7/073055}}.
\newline\urlprefix\url{http://stacks.iop.org/1367-2630/14/i=7/a=073055}

\bibitem{zeitler_linearization_2015}
B.~Zeitler, K.~Floettmann, F.~Grüner,
  \href{https://link.aps.org/doi/10.1103/PhysRevSTAB.18.120102}{Linearization
  of the longitudinal phase space without higher harmonic field}, Phys. Rev. ST
  Accel. Beams 18~(12) (2015) 120102.
\newblock \href {https://doi.org/10.1103/PhysRevSTAB.18.120102}
  {\path{doi:10.1103/PhysRevSTAB.18.120102}}.
\newline\urlprefix\url{https://link.aps.org/doi/10.1103/PhysRevSTAB.18.120102}

\bibitem{luiten_how_2004}
O.~J. Luiten, S.~B. van~der Geer, M.~J. de~Loos, F.~B. Kiewiet, M.~J. van~der
  Wiel, \href{https://link.aps.org/doi/10.1103/PhysRevLett.93.094802}{How to
  {Realize} {Uniform} {Three}-{Dimensional} {Ellipsoidal} {Electron}
  {Bunches}}, Phys. Rev. Lett. 93~(9) (2004) 094802.
\newblock \href {https://doi.org/10.1103/PhysRevLett.93.094802}
  {\path{doi:10.1103/PhysRevLett.93.094802}}.
\newline\urlprefix\url{https://link.aps.org/doi/10.1103/PhysRevLett.93.094802}

\bibitem{schiffrin_optical-field-induced_2013}
A.~Schiffrin, T.~Paasch-Colberg, N.~Karpowicz, V.~Apalkov, D.~Gerster,
  S.~Mühlbrandt, M.~Korbman, J.~Reichert, M.~Schultze, S.~Holzner, J.~V.
  Barth, R.~Kienberger, R.~Ernstorfer, V.~S. Yakovlev, M.~I. Stockman,
  F.~Krausz,
  \href{http://www.nature.com/nature/journal/v493/n7430/full/nature11567.html}{Optical-field-induced
  current in dielectrics}, Nature 493~(7430) (2013) 70--74.
\newblock \href {https://doi.org/10.1038/nature11567}
  {\path{doi:10.1038/nature11567}}.
\newline\urlprefix\url{http://www.nature.com/nature/journal/v493/n7430/full/nature11567.html}

\bibitem{cesar_nonlinear_2017}
D.~Cesar, S.~Custodio, J.~Maxson, P.~Musumeci, X.~Shen, E.~Threlkeld, R.~J.
  England, A.~Hanuka, I.~V. Makasyuk, E.~A. Peralta, K.~P. Wootton, Z.~Wu,
  \href{http://arxiv.org/abs/1707.02364}{Nonlinear response in high-field
  dielectric laser accelerators}, arXiv:1707.02364 [physics]ArXiv: 1707.02364.
\newline\urlprefix\url{http://arxiv.org/abs/1707.02364}

\bibitem{morimoto_diffraction_2018}
Y.~Morimoto, P.~Baum,
  \href{https://www.nature.com/articles/s41567-017-0007-6}{Diffraction and
  microscopy with attosecond electron pulse trains}, Nature Physics 14~(3)
  (2018) 252--256.
\newblock \href {https://doi.org/10.1038/s41567-017-0007-6}
  {\path{doi:10.1038/s41567-017-0007-6}}.
\newline\urlprefix\url{https://www.nature.com/articles/s41567-017-0007-6}

\bibitem{baum_femtosecond_2008}
P.~Baum, A.~Zewail,
  \href{http://www.sciencedirect.com/science/article/pii/S0009261408009925}{Femtosecond
  diffraction with chirped electron pulses}, Chemical Physics Letters 462~(1)
  (2008) 14--17.
\newblock \href {https://doi.org/10.1016/j.cplett.2008.07.072}
  {\path{doi:10.1016/j.cplett.2008.07.072}}.
\newline\urlprefix\url{http://www.sciencedirect.com/science/article/pii/S0009261408009925}

\bibitem{barwick_photon-induced_2009}
B.~Barwick, D.~J. Flannigan, A.~H. Zewail,
  \href{https://www.nature.com/articles/nature08662}{Photon-induced near-field
  electron microscopy}, Nature 462~(7275) (2009) 902--906.
\newblock \href {https://doi.org/10.1038/nature08662}
  {\path{doi:10.1038/nature08662}}.
\newline\urlprefix\url{https://www.nature.com/articles/nature08662}

\bibitem{blaga_imaging_2012}
C.~I. Blaga, J.~Xu, A.~D. DiChiara, E.~Sistrunk, K.~Zhang, P.~Agostini, T.~A.
  Miller, L.~F. DiMauro, C.~D. Lin,
  \href{https://www.nature.com/articles/nature10820}{Imaging ultrafast
  molecular dynamics with laser-induced electron diffraction}, Nature
  483~(7388) (2012) 194--197.
\newblock \href {https://doi.org/10.1038/nature10820}
  {\path{doi:10.1038/nature10820}}.
\newline\urlprefix\url{https://www.nature.com/articles/nature10820}

\end{thebibliography}


\end{document}